\documentclass[conference]{IEEEtran}
\IEEEoverridecommandlockouts
\usepackage{cite}
\usepackage{amsmath,amssymb,amsfonts}
\usepackage{algorithmic}
\usepackage{graphicx}
\usepackage{textcomp}
\usepackage{xcolor}
\usepackage{balance}
\usepackage{dblfloatfix}
\def\BibTeX{{\rm B\kern-.05em{\sc i\kern-.025em b}\kern-.08em
    T\kern-.1667em\lower.7ex\hbox{E}\kern-.125emX}}

\begin{document}

\title{Neuromorphic Processor Employing FPGA Technology with Universal Interconnections}

\author{
\IEEEauthorblockN{Pracheta Harlikar, Abdel-Hameed A. Badawy}
\IEEEauthorblockA{
\textit{Klipsch School of Electrical and Computer Engineering} \\
\textit{New Mexico State University} \\
Las Cruces, New Mexico, USA \\
\{pracheta, badawy\}@nmsu.edu
}
\and
\IEEEauthorblockN{Prasanna Date}
\IEEEauthorblockA{
\textit{Oak Ridge National Laboratory} \\
Oak Ridge, Tennessee, USA \\
datepa@ornl.gov
}
}

\maketitle

\begin{abstract}
Neuromorphic computing, inspired by biological neural systems, holds immense promise for ultra-low-power and real-time inference applications. However, limited access to flexible, open-source platforms continues to hinder widespread adoption and experimentation. In this paper, we present a low-cost neuromorphic processor implemented on a Xilinx Zynq-7000 FPGA platform. The processor supports all-to-all configurable connectivity and employs the leaky integrate-and-fire (LIF) neuron model with customizable parameters such as threshold, synaptic weights, and refractory period. Communication with the host system is handled via a UART interface, enabling runtime reconfiguration without hardware resynthesis. The architecture was validated using benchmark datasets including the Iris classification and MNIST digit recognition tasks. Our post-synthesis results highlight the design’s energy efficiency and scalability, establishing its viability as a research-grade neuromorphic platform that is both accessible and adaptable for real-world spiking neural network applications. This implementation will be released as open-source following project completion.
\end{abstract}


\section{Introduction}
Neuromorphic computing, diverging from the traditional von Neumann architecture, emulates the computational mechanisms of the human brain to achieve exceptional energy efficiency. Studies have demonstrated that neuromorphic systems can deliver energy savings of several orders of magnitude compared to conventional CPU and GPU architectures, making them highly attractive for applications in edge computing, autonomous systems, and the Internet of Things [1], [2]. At the heart of this computing paradigm are spiking neural networks (SNNs), which process information through discrete spikes in a manner similar to biological neurons, offering reduced power consumption and improved real-time performance [3], [4].

Several research institutions have developed neuromorphic hardware platforms to explore biologically inspired computing at scale. Notable examples include Neurogrid from Stanford, a mixed-analog-digital multichip system enabling large-scale simulations of cortical networks [5], and Braindrop, a mixed-signal architecture developed to model dynamical systems efficiently [6]. NeurRAM introduced compute-in-memory functionality for versatile AI inference tasks [7], while BrainScaleS from Heidelberg University emphasizes verification and accelerated analog computation [8]. Projects like SpiNNaker from the University of Manchester have demonstrated massively parallel digital architectures with over a million cores to support real-time SNN simulations [9], and DARWIN, developed at Zhejiang University, presents a neuromorphic co-processor optimized for low-latency spike processing [10]. Meanwhile, commercial research efforts such as Intel’s Loihi [11] and IBM’s TrueNorth [12] have pushed on-chip learning and programmable neurosynaptic arrays forward. Other implementations like DANNA and DANNA 2 [13], [14], and compact FPGA-based systems [15],[16] further illustrate the wide spectrum of scalable, event-driven, and low-power neuromorphic solutions being explored.

While these advancements represent significant strides in neuromorphic computing, many of the existing platforms remain proprietary or lack open-source accessibility, which poses challenges for wider academic and independent research adoption. The lack of freely available hardware and design flexibility makes it difficult for smaller research groups or educational institutions to build upon or tailor these systems to their specific application needs.

Despite their promise, neuromorphic systems face several practical barriers, including limited hardware availability, underdeveloped toolchains, and high system complexity, which restrict broader adoption and experimentation [17]. To address these limitations, FPGA-based implementations have emerged as viable solutions due to their reconfigurability, parallelism, and lower development costs [18]. In this work, we propose a low-cost, open-source neuromorphic processor implemented on a Xilinx Zynq-7000 FPGA platform. The processor supports all-to-all connectivity among neurons, customizable synaptic weights, thresholds, and refractory periods using a leaky integrate-and-fire (LIF) model. Communication with the system is achieved using a Universal Asynchronous Receiver-Transmitter (UART) interface, allowing runtime parameter reconfiguration without requiring hardware re-synthesis.

This paper presents the design, implementation, and validation of this processor using benchmark datasets including the Iris dataset and the MNIST image classification task. This system demonstrates its computational correctness, energy efficiency, and potential as a scalable neuromorphic research platform. Furthermore, the architecture's modular design and support for real-time interfacing make it well-suited for future extensions involving embedded processing and online learning mechanisms.

\section{Design \& Implementation}

\subsection{Neuromorphic Model}

This implementation utilizes leaky integrate-and-fire (LIF) neurons, which accumulate charge from incoming synapses and emit a spike once the membrane potential reaches a defined threshold, after which they reset to a specified state. The neuron model supports all essential parameters: inputs, thresholds, refractory periods, membrane potential, leak, and outputs. Synaptic parameters include integer weights (0–255) and configurable delay values (default 1 clock cycle, adjustable up to 255). The architecture is fully parametric and user-configurable, allowing flexibility to increase the 8-bit input width and the number of inputs to suit various application requirements. 

\subsubsection{{Discrete-Time Leaky Integrate-and-Fire (LIF) Model}}
To formally characterize the dynamics of the implemented neuron, we adopt the discrete-time leaky integrate-and-fire (LIF) model (Euler Update). The update rule describes how the membrane potential evolves at each clock cycle under the influence of weighted input spikes, leakage, and thresholding. The equations below summarize the membrane potential integration, spike generation, reset, and refractory behavior used in our hardware realization. 

Let $k$ index clock cycles (sampling period $\Delta t$). The membrane update is given by:
\begin{equation}
\tilde{v}[k+1] = \left(1 - \frac{\Delta t}{\tau_m}\right)v[k] + \frac{\Delta t}{C_m}\left(\sum_j w_j s_j[k] + I_{\text{bias}}[k]\right)
\end{equation}

with thresholding, reset, and refractory:
\begin{align}
y[k+1] &= 
\begin{cases}
1, & \text{if } \tilde{v}[k+1] \geq V_{\text{th}} \text{ and } r[k] = 0 \\
0, & \text{otherwise}
\end{cases} \\
v[k+1] &= 
\begin{cases}
0, & \text{if } y[k+1] = 1 \text{ or } r[k] > 0 \\
\tilde{v}[k+1], & \text{otherwise}
\end{cases} \\
r[k+1] &= 
\begin{cases}
R_{\text{ref}}, & \text{if } y[k+1] = 1 \\
\max(0,\, r[k]-1), & \text{otherwise}
\end{cases}
\end{align}

\noindent\textbf{where:}
\begin{tabular}{ll}
$v[k]$ & membrane potential \\
$\tilde{v}[k+1]$ & pre-threshold update \\
$y[k]\in\{0,1\}$ & output spike \\
$\tau_m$ & membrane time constant \\
$C_m$ & membrane capacitance \\
$w_j$ & synaptic weight for input $j$ \\
$s_j[k]\in\{0,1\}$ & input spike at cycle $k$ \\
$I_{\text{bias}}[k]$ & bias/tonic input \\
$V_{\text{th}}$ & threshold \\
$r[k]$ & refractory counter (cycles) \\
$R_{\text{ref}}$ & refractory length (cycles) \\
\end{tabular}

\subsubsection{{Hardware (Fixed-Leak) Realization}}

If the leak is implemented as a fixed decrement per cycle (when the neuron is active), use:
\begin{equation}
\tilde{v}[k+1] = v[k] + \sum_j w_j s_j[k] - \lambda \cdot \mathbf{1}\{v[k] \neq 0\}
\end{equation}

followed by the same threshold/reset/refractory equations above, where $\lambda$ is the per-cycle leak step and $\mathbf{1}\{\cdot\}$ is the indicator function.

\subsection{Hardware}

The ZedBoard development platform [17] was used as the primary hardware interface for this study. It features a Xilinx Zynq-7000 SoC, which integrates both programmable logic and a processing system. The board was programmed and configured using the Xilinx Vivado Design Suite [18]. For serial communication, the RXD (Receive Data) and TXD (Transmit Data) lines were used to implement the UART protocol at a baud rate of 9600. The system functioned using the default clock frequency of 100 MHz. 

\subsection{Processor Architecture}\label{AA}
\begin{figure}[h!]
\centering
\includegraphics[width=\linewidth]{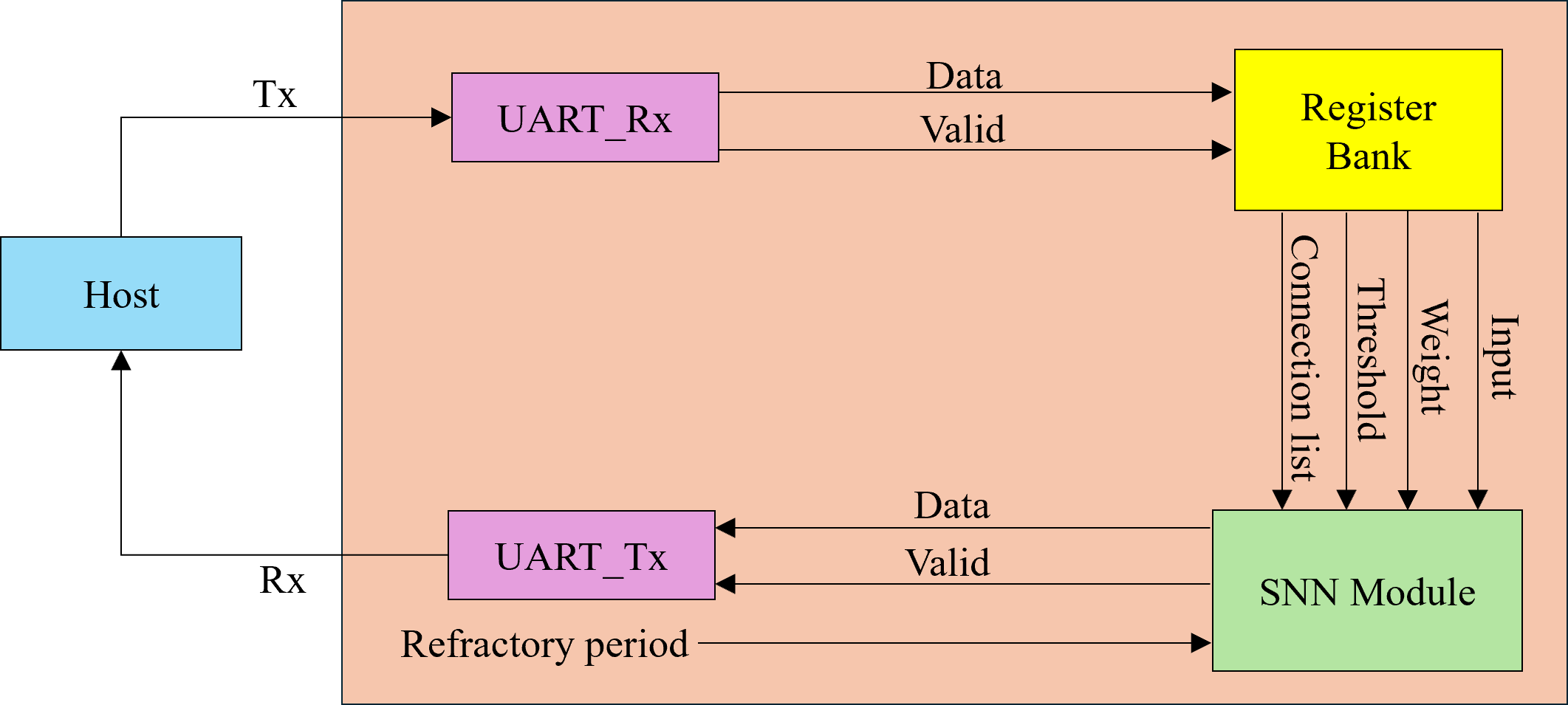}
\caption{Processor Architecture}
\end{figure}

The processor architecture, illustrated in Fig. 1, presents a modular design tailored for the implementation of a Spiking Neural Network (SNN) on an FPGA platform. The system facilitates communication with an external host via a Universal Asynchronous Receiver/Transmitter (UART) interface and comprises three key components: the UART modules, a register bank, and the core SNN processing module.

The UART receiver module (UART\_Rx) is tasked with decoding serial input data from the host and converting it into parallel data. This incoming data primarily contains SNN configuration parameters such as neuron weights, input spikes, thresholds, connection lists, and refractory periods. Each byte (8 bits) of data received is validated and subsequently routed to the register bank, where it is stored in appropriate control and configuration registers. The register bank acts as a temporary storage and routing interface, parsing incoming data and organizing it into structured fields that define the behavior of individual neurons within the SNN module. 

Upon successful register updates, parameters such as inputs, weights, threshold and connection list are passed from the register bank to the SNN module, which represents the core computation block. The SNN module is composed of an array of homogeneous neurons, each capable of independently processing input spikes based on their internal parameters. Neurons compute membrane potentials and generate output spikes based on the defined thresholds and refractory dynamics.

In the reverse direction, output data generated by the SNN module—such as spike events or updated refractory counters—is transferred back to the host system. This is achieved via the UART transmitter module (UART\_Tx). The UART\_Tx collects parallel data from the SNN module and converts it into serial format, ensuring compatibility with standard UART protocols at a baud rate of 9600. Transmission is gated by a validation signal to ensure data integrity.

The system operates synchronously using a 100 MHz system clock and is fully configurable at runtime through the UART interface, enabling dynamic reconfiguration of SNN parameters without the need for FPGA reprogramming. This architecture provides a flexible and efficient platform for implementing real-time spiking neural networks, making it suitable for low-latency neuromorphic computing applications on FPGA hardware such as the ZedBoard.

The processor can be readily scaled from one dataset to another with minimal design effort. By reprogramming the registers, the interconnections can be reconfigured, and by updating the threshold weights and related parameters, the processor can be adapted for new applications. Once reprogrammed, each neuron requires 2 clock cycles to process its input data; thus, each layer in the spiking neural network (SNN) incurs a 2-cycle latency. The total end-to-end inference latency, measured from the arrival of the input spike to the generation of the corresponding output spike, is 5 clock cycles: 1 cycle for input sampling, 2 cycles for input-layer processing, and 2 cycles for output-layer processing. Since both the Iris and MNIST networks employ the same two-layer structure (input and output), the total latency is identical—5 cycles—for both cases.

\subsection{Neuron connectivity}

The processor architecture leverages a scalable, fully connected neuron topology where each neuron in the SNN module can communicate with every other neuron through a multiplexer-based connection scheme. This interconnect logic is illustrated in Fig. 2 and Fig. 3 below.

\begin{figure}[htbp]
\centerline{\includegraphics[width=1\linewidth]{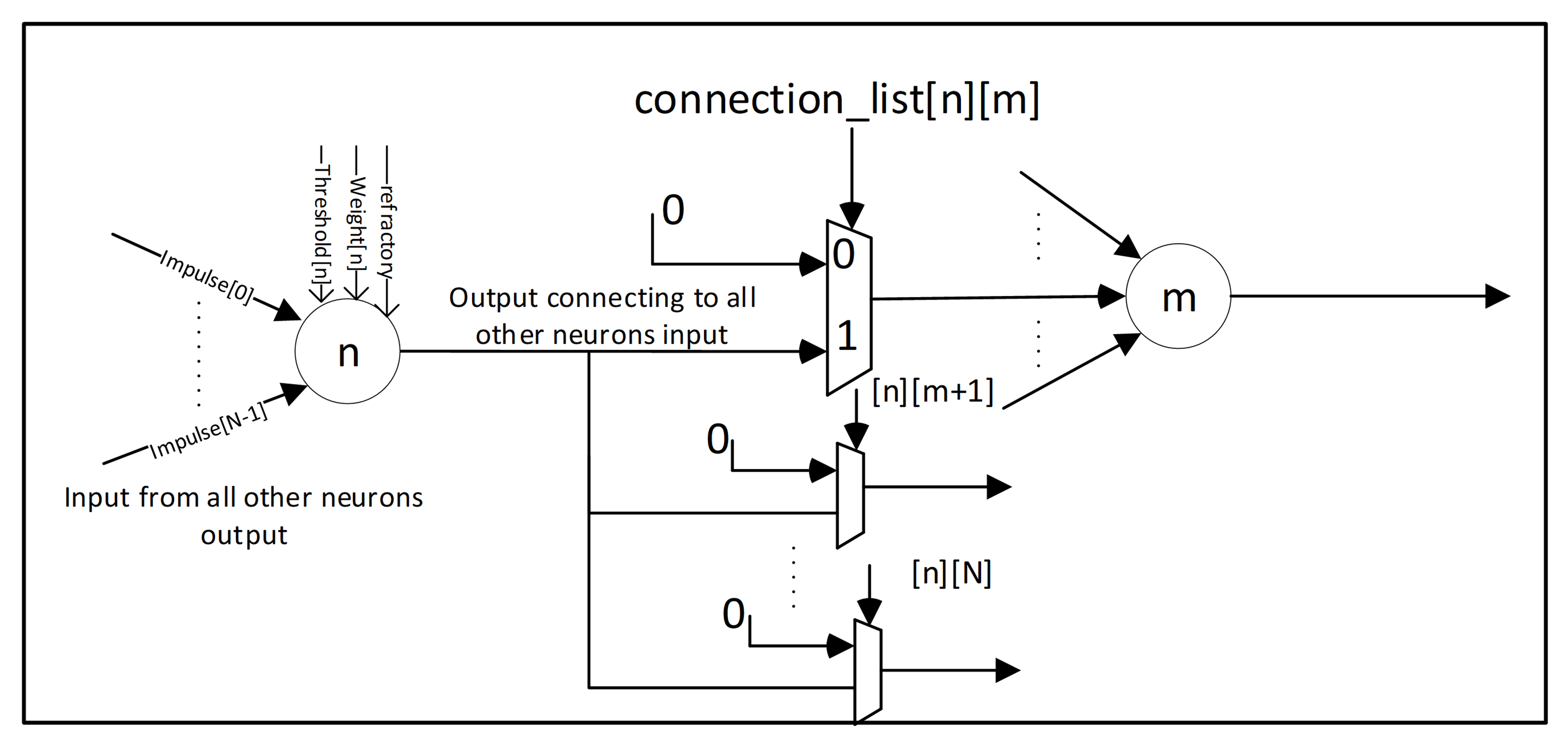}}
\caption{Architecture for All-to-All Neuron Connectivity with Selective Routing}
\label{fig}
\end{figure}

As shown in Fig. 2, neuron n receives inputs from all other N neurons and uses internally stored parameters such as weight and threshold to determine its membrane potential. Once the potential reaches the threshold, the neuron emits a spike and resets. After a spike is generated, the neuron does not process any new input spikes during its programmed refractory period. The outgoing spike is routed to all other neurons in the network using a matrix of multiplexers. The connection list[n][m] configuration parameter governs the connectivity between neurons n and m. If the value at connection list[n][m] is set to 1, the corresponding multiplexer enables the connection, forwarding the output of neuron n to neuron m's input. Otherwise, the multiplexer routes a zero, effectively disabling the connection. This selective enablement ensures efficient dynamic configurability without hardware redundancy. 

\begin{figure}[htbp]
\centerline{\includegraphics[width=1\linewidth]{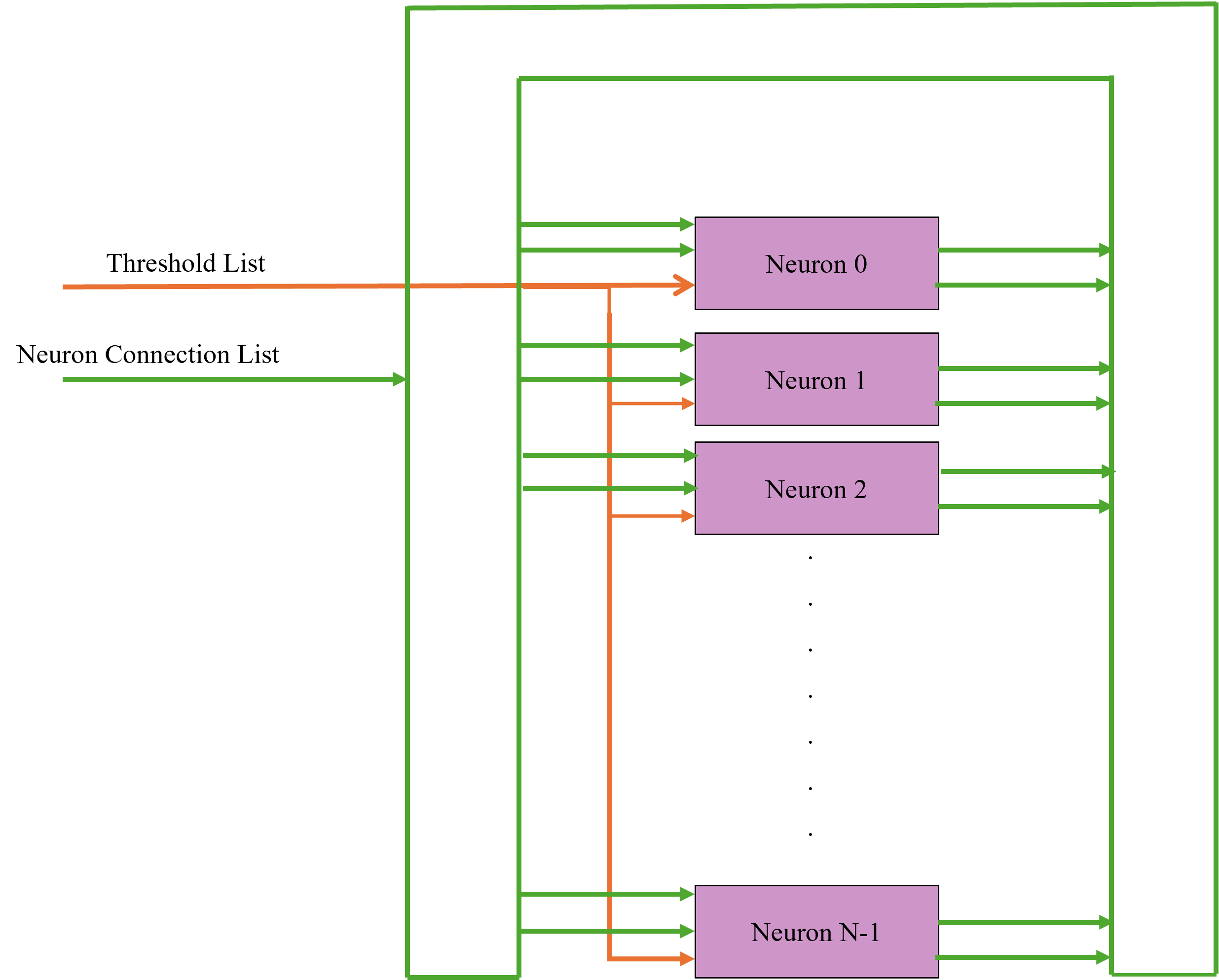}}
\caption{Block Diagram of Parametric SNN Neuron Network}
\label{fig}
\end{figure}

Fig. 3 depicts a block diagram of the SNN neuron array, where each neuron labeled from 0 to N-1 receives threshold values and connectivity patterns from a centralized threshold list and neuron connection list, respectively. These control parameters are stored in the register bank and are configured through the UART interface. This design approach enables parallel execution, with each neuron operating independently based on its local parameters while also contributing to the collective network behavior.

This mux-based all-to-all connectivity not only supports scalable network topologies but also enables runtime configurability of the synaptic connections, allowing the system to adapt to different SNN architectures and datasets without requiring re-synthesis. Such a modular design enhances the processor’s applicability to a wide range of neuromorphic computing tasks, supporting both dense and sparse connectivity by simply modifying the connection list.

The resource utilization of each neuron is directly dependent on the number of inputs. This scalability is handled during FPGA compilation, where the design dynamically adjusts its resource requirements according to the specified input configuration. Table I below describes the resource utilization per neuron.

\begin{table}[htbp]
\caption{Utilization Report per neuron}
\centering
\resizebox{\columnwidth}{!}{%
\begin{tabular}{|c|c|c|}
\hline
\textbf{Total number of neurons}& \textbf{Slice-LUTs} & \textbf{Slice Reg.} \\
\hline
8& 12& 13\\
\hline
74& 24& 13\\\hline
\end{tabular}%
}
\label{tab:iris_util}
\end{table}

 \section{Testing \& Results}

This implementation was tested on the ZedBoard FPGA platform and successfully simulated for two benchmark datasets: the Iris dataset and the MNIST image classification task. For each test case, the output from the Zynq-7000 SoC was monitored via the console, and a post-bitstream generation utilization report was obtained using the Vivado Design Suite to analyze hardware resource usage. 
 
\subsection{Iris dataset}
\begin{figure}[htbp]
\centerline{\includegraphics[width=1\linewidth]{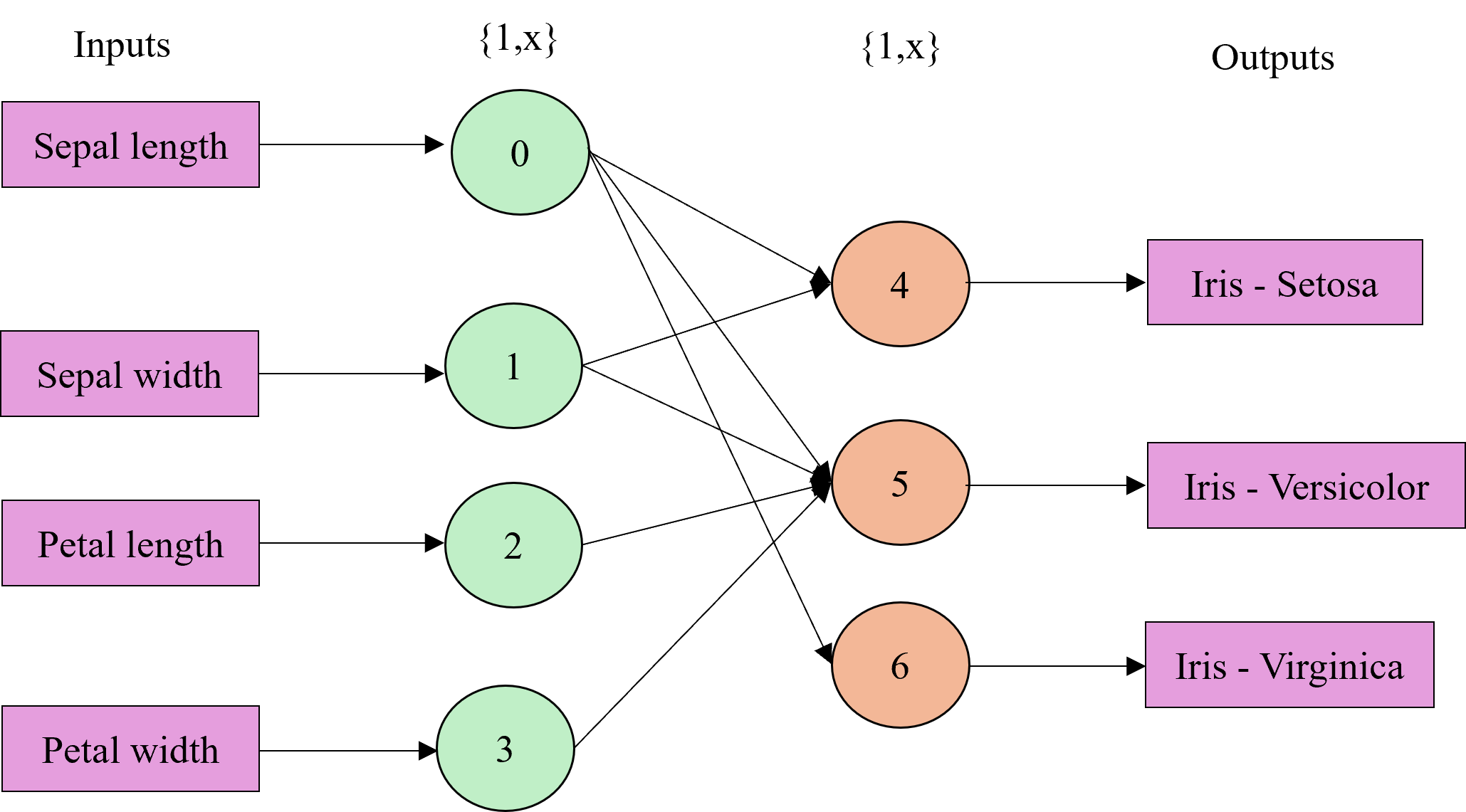}}
\caption{Iris dataset architecture}
\label{fig}
\end{figure}

For evaluating the proposed SNN architecture, the Iris dataset was utilized as a benchmark. The network configuration employed for this task is depicted in Fig. 4. It consists of two layers: an input layer with four neurons and an output layer with three neurons. Each input neuron corresponds to one of the four numerical features in the dataset—sepal length, sepal width, petal length, and petal width.

The network processes 50 samples per class, with a total of 150 input samples (50 samples × 3 flower types), each containing four feature values. For each sample, four parallel spike trains are generated at the input layer. The spiking neuron model used is the leaky integrate-and-fire (LIF) model, with a threshold set to 1 and a refractory period of 2 clock cycles (in the figure mentioned as x in curly bracket). Neurons in the output layer (neurons 4, 5, and 6) are assigned to identify Iris-setosa, Iris-versicolor, and Iris-virginica, respectively. Upon processing, only one of the output neurons spikes to indicate the classification result for each input instance, enabling clear and interpretable decision-making. 

After successful simulation, the design was synthesized and implemented on the ZedBoard using the Vivado Design Suite. The post-implementation utilization report is summarized in Table II. The total on-chip power consumption was recorded as 0.113 W, with a junction temperature of 26.3°C, demonstrating both energy efficiency and thermal stability of the hardware implementation. 

\begin{table}[htbp]
\caption{Utilization Report of Iris Dataset}
\centering
\resizebox{\columnwidth}{!}{%
\begin{tabular}{|l|c|c|c|c|c|}
\hline
\textbf{Name} & \textbf{Slice-LUTs} & \textbf{Slice Reg.} & \textbf{F7 Muxes} & \textbf{IO} & \textbf{BUFG} \\
\hline
snn\_top          & 741 & 608 & 61 & 18 & 2 \\
\hline
u\_ascii\_to\_hex & 1   & 9   & 0  & 0  & 0 \\
\hline
u\_clk\_div       & 44  & 33  & 0  & 0  & 0 \\
\hline
u\_rx            & 38  & 32  & 0  & 0  & 0 \\
\hline
u\_snn\_proc      & 522 & 408 & 40 & 0  & 0 \\
\hline
u\_tx            & 134 & 124 & 21 & 0  & 0 \\
\hline
\end{tabular}%
}
\label{tab:iris_util}
\end{table}

\begin{figure}[htbp]
\centerline{\includegraphics[width=1\linewidth]{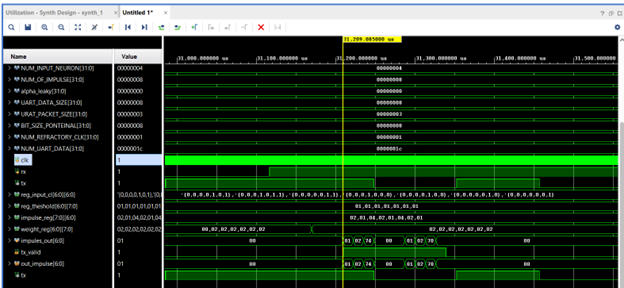}}
\caption{Simulation Waveform Analysis for Iris Dataset}
\label{fig}
\end{figure}

Fig. 5 presents the post-synthesis simulation waveform captured in the Vivado Design Suite, validating the correct functionality of the implemented SNN architecture during inference on the \textit{Iris} dataset. The simulation captures the behavior of key signals during one input transaction cycle.

The testbench initializes control parameters as follows:

\begin{itemize}
    \item \verb|NUM_INPUT_NEURON = 4| – representing four feature inputs
    \item \verb|NUM_OF_IMPULSE = 4| – one impulse per feature
    \item \verb|UART_DATA_SIZE = 8 bits|
    \item \verb|NUM_UART_DATA = 0x1C| – 28 bytes transferred via UART
\end{itemize}
During simulation, the \verb|rx| signal receives serial data from the host system, encoded according to the UART protocol. This data contains the threshold values, weight matrix, and input spike patterns for an \textit{Iris} dataset sample. As data is received, it populates internal registers:

\begin{itemize}
    \item \verb|reg_input_clf|, \verb|reg_threshclf|, and \verb|impulse_reg| – holding the spike input pattern and neuron parameters
    \item \verb|weight_reg| – storing synaptic weights
\end{itemize}
At timestamp 31,299.085 µs, valid impulses are registered across the four input neurons, as indicated by the \verb|impulse_reg| values (e.g., \verb|01, 01, 04, 02|), which correspond to quantized petal width inputs for different samples. These spike inputs trigger weighted contributions to connected output neurons.

The \verb|out_impulse| signal shows the final spiking response from the output neurons. For example, the sequence \verb|01, 02, 70| indicates activity at specific output neurons corresponding to a classification result (e.g., neuron 4 spiking for \textit{Setosa}).

The \verb|tx_valid| signal flags the transmission of classification results via the \verb|tx| line, confirming correct back-propagation of results to the host system. The complete classification cycle is successfully completed within microseconds, demonstrating the efficiency and correctness of the implemented pipeline.

This waveform verifies that the architecture correctly decodes UART input, updates neuron states, processes spikes, and outputs the predicted class label based on the input features.

\subsection{MNIST dataset}
\begin{figure}[htbp]
\centerline{\includegraphics[width=1\linewidth]{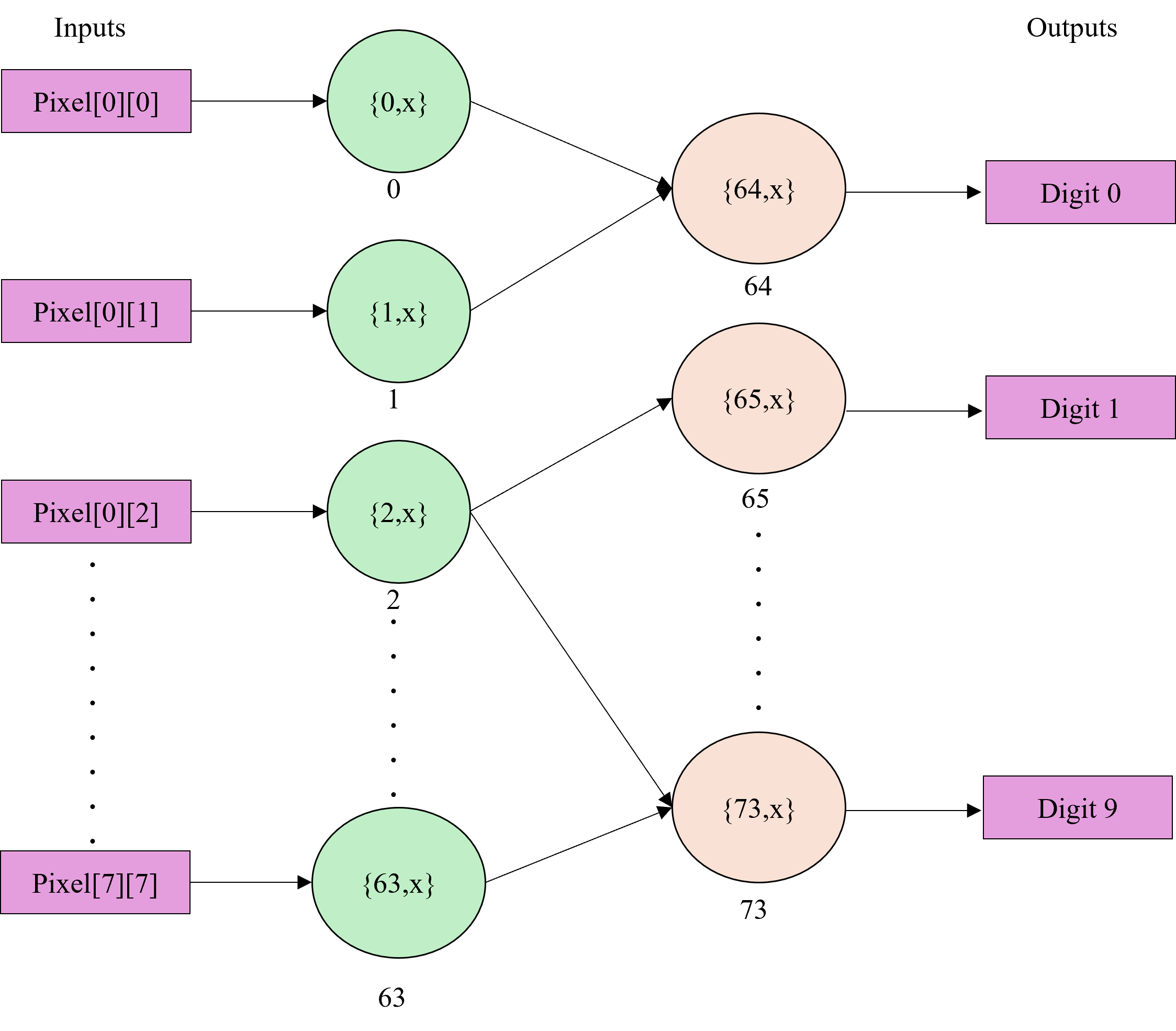}}
\caption{MNIST dataset architecture}
\label{fig}
\end{figure}

Fig. 6 represents a two-layer Spiking Neural Network (SNN) architecture developed and implemented on FPGA for digit classification using the MNIST dataset, resized to 8×8 grayscale images. Each image was first converted to grayscale and then binarized through thresholding, where pixels above the threshold were encoded as spike inputs (represented by ‘1’), while the rest remained silent (‘0’). These binary inputs were mapped to 0 to 63 input neurons, each corresponding to one pixel in the image. The spikes were transmitted in parallel, with a refractory period of four clock cycles (represented as x in the above architecture) to prevent temporal overlap. The output layer consisted of 10 neurons, each representing a digit class from 0 to 9 mapped from neurons 64 to 73. 

A digit was classified based on the neuron with the highest accumulated activation. The entire pipeline, including image processing and spike encoding, was implemented in Python and interfaced with the FPGA via UART. On the FPGA, the design comprised several modules including the SNN processing unit, UART receiver and transmitter, and a clock division module. Synaptic weights, threshold levels, and neuron connectivity lists were transmitted to the FPGA alongside the spike inputs. The post-synthesis resource utilization report, summarized in Table III, indicates that the SNN processing module (u\_snn\_proc) constituted the primary share of hardware usage, utilizing 12,774 slice LUTs and 7,603 registers. In contrast, auxiliary modules handling communication and control—such as u\_rx, u\_tx, and u\_clk\_div—consumed significantly fewer resources. The system correctly tested all digit classes (0–9), highlighting the efficiency of the spike-based neural processing framework within a resource-constrained FPGA environment. Additionally, the total on-chip power consumption was measured at 0.179 W, with the junction temperature stabilizing at 27.1°C, confirming the design's low-power and thermally stable operation. 

We chose 9600 baud UART interface as this baud rate the transmission bandwidth is limited, resulting in significant latency for parameter updates. Specifically, for a 74-neuron system, the total number of UART transactions required is 898 cycles. Given that each UART transaction requires 104.17 µs, the complete register update requires approximately 93.54 ms.

The transaction breakdown is as follows:

\begin{itemize}
    \item CL registers (74 × 74): Each CL requires 10 transactions, yielding a total of 740 transactions.
    \item Threshold registers (74 × 8): A total of 74 transactions.
    \item Weight registers (74 × 8): A total of 74 transactions.
    \item Impulse registers (74 bits): Represented as 64 bits + 16 bits (10 used + 6 unused), requiring 10 transactions.
\end{itemize}

Thus, the total is 898 UART transactions. For comparison, a single-neuron system requires only four transactions, which at the same baud rate amounts to 416.68 µs. 

\begin{table}[htbp]
\caption{Utilization Report of MNIST Dataset}
\centering
\resizebox{\columnwidth}{!}{%
\begin{tabular}{|l|c|c|c|c|c|}
\hline
\textbf{Name} & \textbf{Slice-LUTs} & \textbf{Slice Reg.} & \textbf{F7 Muxes} & \textbf{F8 Muxes}& \textbf{IO}\\
\hline
snn\_top          & 12989& 7849& 722& 216& 18\\
\hline
u\_ascii\_to\_hex & 0& 11& 0  & 0  & 0 \\
\hline
u\_clk\_div       & 49& 33  & 0  & 0  & 0 \\
\hline
u\_rx            & 40& 32  & 0  & 0  & 0 \\
\hline
u\_snn\_proc      & 12774& 7603& 705& 208& 0 \\
\hline
u\_tx            & 124& 165& 17& 8& 0 \\
\hline
\end{tabular}%
}
\label{tab:iris_util}
\end{table}

\begin{figure}[htbp]
\centerline{\includegraphics[width=1\linewidth]{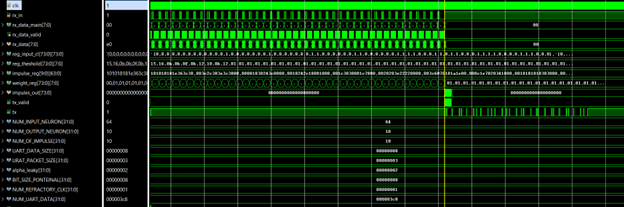}}
\vspace{-2mm}
\caption{Simulation Waveform Analysis for MNIST Dataset}
\label{fig}
\end{figure}

Fig. 7 shows the simulation waveform captured using Xilinx Vivado for the FPGA implementation of the MNIST 8×8 image classification using a two-layer SNN. The signal \verb|clk| represents the main clock driving the system, and \verb|rx_data_main| along with \verb|rx_data_valid| confirms successful serial communication over UART. The spike inputs are visible on the \verb|rx_input| bus, representing the thresholded 8×8 binary image, where ‘1’ indicates an active pixel and generates a corresponding spike impulse. The refractory period is configured using the \verb|rep_threshold| signal, set to four clock cycles to prevent overlapping activations. Synaptic weights (\verb|weight_reg|) and membrane potentials (\verb|impulse_reg|) are initialized and updated during the inference phase. The spike impulses processed and propagated through the SNN are visualized on the \verb|impulses_out| signal, which shows active spikes being generated in sync with the input spike train. The signals \verb|tx| and \verb|tx_valid| indicate outgoing UART transmissions carrying processed classification results. Parameter values for the simulation, including the number of input neurons (\verb|NUM_INPUT_NEURON| = 64), output neurons (\verb|NUM_OUTPUT_NEURON| = 10), and refractory cycles (\verb|NUM_REFRACTORY_CLK| = 4), are clearly initialized and can be seen latched in the waveform. Additionally, meta-parameters like \verb|alpha_leaky|, spike potential resolution, and UART packet configurations are set during initialization. The waveform validates correct impulse generation, neuron state transitions, and communication sequences, thus confirming the functional accuracy of the hardware implementation in processing the MNIST dataset. 

\section{Experimental Setup}

\begin{figure}[htbp]
\centerline{\includegraphics[width=0.75\linewidth]{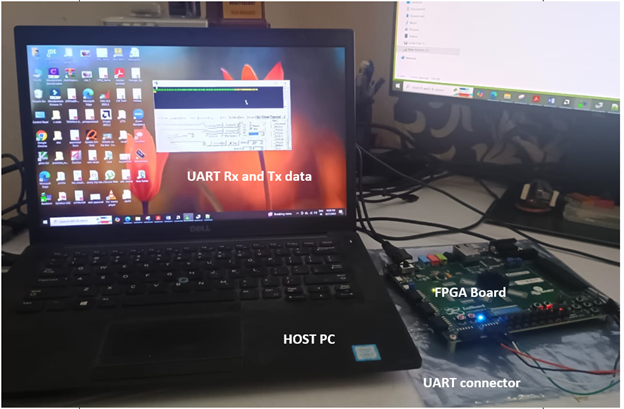}}
\vspace{-2mm}
\caption{Experimental Setup}
\label{fig}
\end{figure}
Fig. 8 presents the experimental hardware-software co-simulation setup used for evaluating the FPGA-based implementation of spiking neural networks (SNNs) across multiple datasets, including MNIST and Iris. The setup comprises a host PC and an FPGA development board connected via a UART serial interface. The host PC is responsible for dataset preprocessing, neural encoding, and communication with the FPGA. For image-based datasets like MNIST, input samples are resized and binarized to convert them into spike trains. For tabular datasets such as Iris, feature values are normalized and encoded into spike patterns or binary activation vectors. A Python-based script handles all preprocessing and UART communication.

Once the data is encoded, it is transmitted serially from the host PC to the FPGA board through UART. Along with the input spikes, associated neural network parameters—such as weights, thresholds, refractory periods, and configuration metadata—are also sent to the FPGA. The board hosts a custom Verilog-based SNN processing unit that receives this data and performs inference in hardware, simulating biologically inspired spike-based computation across layers of neurons.

Upon processing, the FPGA generates output spikes corresponding to the predicted class label. This output is transmitted back to the host PC over the same UART interface. The host application decodes the results and logs them for visualization and performance analysis. As shown in the figure, a terminal interface on the host PC captures both the transmitted inputs and received outputs in real time, verifying system functionality and enabling debugging. This platform-agnostic experimental framework provides flexibility to evaluate SNN models on various datasets while utilizing FPGA acceleration for energy-efficient neural processing.

\section{Conclusion}\label{SCM}
This work demonstrates a functional FPGA-based neuromorphic processor capable of real-time inference using spiking neural networks with all-to-all configurable connectivity. The architecture employs the leaky integrate-and-fire model and supports essential neuron and synapse parameters including threshold, refractory period, and synaptic weights. Comprehensive testing was carried out using the Iris and MNIST datasets, correctly classifying all test cases with low power consumption and resource usage. The processor is fully reconfigurable via UART, enabling flexibility for a wide range of applications without requiring FPGA re-synthesis. While the current implementation validates the design's effectiveness and scalability.

\section{Future work}

The current design is limited by UART-based communication, which introduces programming delays of nearly 100 ms for moderate-sized systems. This bottleneck can be alleviated by employing higher-speed protocols such as Ethernet or USB, both supported on Zynq and ZedBoard platforms, enabling faster reconfiguration and improved real-time performance. Future directions include integrating the neuromorphic processor with embedded processors running real-time operating systems to improve communication flexibility, and implementing and testing the design on FPGA hardware with real-time datasets. Additional research will focus on ASIC implementation for performance comparison with conventional processors and exploring on-chip learning mechanisms such as Spike-Timing-Dependent Plasticity (STDP) on FPGA platforms.

\balance

\end{document}